\documentclass{raa}            

\usepackage{graphicx,times}  
\usepackage{natbib}
\usepackage{booktabs}
\usepackage{amssymb,amsmath}
\bibpunct{(}{)}{;}{a}{}{,}
\usepackage{lineno}
\usepackage{CJK}
\usepackage[pagebackref=true,colorlinks,linkcolor=blue,anchorcolor=blue,citecolor=blue]{hyperref}

\begin{document}

  \title{Homogenization of the Stetson Photometry with the BEST Database}

   \volnopage{Vol.0 (20xx) No.0, 000--000}    
   \setcounter{page}{1}   

   \author{Zhirui Li \inst{1,2}
      \and Bowen Huang \inst{3, 4}
      \and Kai Xiao \inst{2, 3}
      \and Haibo Yuan \inst{3, 4}
      \and Yang Huang \inst{2, 1, 3}
      \and Dongwei Fan \inst{1, 2, 5}
      \and Mingyang Ma \inst{3, 4}
      \and Tao Wang \inst{3, 4}
   }

   \institute{National Astronomical Observatories, Chinese Academy of Sciences, Beijing 100101, China;
   \and
   School of Astronomy and Space Science, University of Chinese Academy of Sciences, Beijing 100049, China;
    \and
    Institute for Frontiers in Astronomy and Astrophysics, Beijing Normal University, Beijing, 102206, China;
    \and
    School of Physics and Astronomy, Beijing Normal University No.19, Xinjiekouwai St, Haidian District, Beijing, 100875, China;
    \and
    National Astronomical Data Center, Beijing 100101, China.
      \\
    Corresponding author: Kai Xiao (xiaokai@ucas.ac.cn)\\
\vs\no
   {\small Accepted 19 February 2026}}

\abstract{As one of the most widely recognized high-quality standard stars, the Stetson standards have been extensively used as a photometric reference for calibrating other surveys. In this work, we present an independent validation and re-calibration of the Stetson standard star photometry using the BEST database. Based on typically 30,000-70,000 calibration stars per band, we find that the original Stetson photometry achieves field-to-field zero-point precisions of approximately 10--40\,mmag in the $UBVRI$-band. In addition, significant spatially dependent magnitude offsets are detected within individual Stetson fields for all bands, with magnitudes exceeding 1\%, probably caused by the calibration errors in the Stetson photometry. After correcting those systematic errors, the agreement between the Stetson and BEST photometry is improved to $\sim$5\,mmag for individual fields for $BVRI$-band. The re-calibrated photometry is further validated using the SCR standards, yielding agreement better than 10\,mmag for individual stars in the $BVRI$ bands and confirming zero-point precisions of 2--4\,mmag in the $BVI$ band. The precisions is further confirmed by checks using Gaia DR3 broad band colors. These results highlight the power of the BEST database for improving photometric calibration and suggest that, if feasible, it be incorporated into the calibration process of future releases of the Stetson standard catalog.
}

   \authorrunning{Li et al.}       
   \titlerunning{Photometric Homogenization of the Stetson Photometry}  

   \maketitle

\section{Introduction}
The luminosity of astronomical objects is directly related to their fundamental physical properties, making high-precision photometry essential for astronomical observations. 
Accurate reference catalogs enable reliable photometric measurements by allowing observers to calibrate target sources against standard stars, and therefore play a central role in photometric calibration.

The Stetson standard star database \citep{Stetson2000, Stetson2005, Stetson2019}, first released in the spring of 2000 \citep{Stetson2000} and continuously updated\footnote{\href{https://www.canfar.net/storage/list/STETSON/}{https://www.canfar.net/storage/list/STETSON/}}, was compiled from multiple observations obtained with a variety of CCD-based telescopes (e.g., the Kitt Peak 4\,m, Cerro Tololo 4\,m in the 2000 release, and the WIYN 3.5\,m telescopes) and provides a photometric standard star catalog in the Johnson--Kron--Cousins \textcolor{black}{(JKC)} $UBVRI$ system \citep{JohnsonUBV, Johnson1963, KronRI} calibrated against the Landolt standards \citep{Landolt1973, Landolt1983, Landolt1992}. As one of the most widely recognized high-quality standard stars, the Stetson standards have been extensively employed for photometric calibration and validation in many major photometric projects, including the Sloan Digital Sky Survey (SDSS; \citealp{SDSS2005, Ivezic2007}), the Hubble Space Telescope Advanced Camera for Surveys (ACS; \citealp{Sarajedini2007}), the Swift ultraviolet/optical telescope (UVOT; \citealp{uvot}), and Gaia early data release 3 \citep{GaiaSSSS2012, GaiaEDR2021}.

Even today, the Stetson standard stars remain highly valuable owing to their multi-band coverage and high signal-to-noise ratios, particularly for faint sources, making them well suited for the photometric calibration of surveys targeting faint objects \textcolor{black}{(e.g., useful for the photometric calibration of the Chinese Space Station Telescope; \citealt{2018cosp...42E3821Z})}. Nevertheless, several factors may introduce non-negligible systematic errors into the Stetson photometry. First, the calibration is tied to the Landolt standards, which themselves exhibit systematic uncertainties at $\sim 1\%$ level \citep{Huang2026}. Second, corrections to the stellar flat-field were not fully accounted for. 
To fully exploit the power of the Stetson standards, it is therefore necessary to carry out both calibration validation and high-precision photometric re-calibration of the Stetson photometry.

The Best Star (BEST) database\textbf{\footnote{\href{https://doi.org/10.12149/101651}{https://doi.org/10.12149/101651}; \href{https://nadc.china-vo.org/data/best/}{https://nadc.china-vo.org/data/best/}}} is a recently constructed standard star catalogs developed by \cite{BEST}. Its first version includes i) more than 200 million high-precision standard stars distributed across the entire sky in over 200 photometric bands (e.g., Landolt $UBVRI$-band), ii) billions of accurately calibrated photometric measurements, and iii) all-sky Pan-STARRS $grizy$ photometry \citep{BEST}. The multi-band standard star catalog was constructed using the stellar color regression (SCR) method and the corrected Gaia BP/RP (XP) based synthetic photometry (XPSP) method \citep{Xiao2023}. The SCR method was first proposed by \cite{Yuan2015} and has been widely applied to the photometric calibration of wide-field photometric surveys, including the re-calibration of the SDSS Stripe 82 magnitudes \citep{Huang2022} and colors \citep{Yuan2015}, the photometric re-calibration of SkyMapper DR2 \citep{Huang2021}, Gaia EDR3 magnitudes \citep{Yang2021} and colors \citep{Niu2021}, and Pan-STARRS1 \citep{Xiao2022,2023ApJS..268...53X}, as well as the photometric calibration of the SAGES $uv$ \citep{Fan2023}, $gri$ \citep{Xiao2023tongbao} and DDO51 (K. Xiao et al. in preparation) bands. The XPSP method was developed based on the XP slitless spectra of over 200 million stars provided by Gaia DR3 and was discussed in \cite{GaiaSP2023}. Building upon corrections for magnitude-, color-, and extinction-dependent systematics in Gaia DR3 slitless spectra \citep{Huang2024}, this method was further improved by \cite{Xiao2023} and successfully applied to the calibration of surveys such as J-PLUS DR3 \citep{Xiao2023} and S-PLUS DR4 \citep{Xiao2024} photometry. The capability of the BEST database has been demonstrated in a number of recent studies. For example, \cite{Li2025} re-calibrated S-PLUS USS data to millimagnitude (mmag) precision, \cite{Huang2026}~validated and re-calibrated Landolt standard stars, \cite{Ma2025}~achieved high-precision photometric calibration of Chinese digitized photographic plates, \cite{2025ApJ...982L..27X,Xiao2025MST} and K. Xiao et al. (in preparation) calibrated data from CMOS-based Mini-Sitian and Sitian prototype telescopes to mmag precision, respectively, and \cite{Rui2025Tianyu}~conducted comparative analyses of data from the Tianyu project. The BEST database provides high-quality data for independent calibration validation and re-calibration of the Stetson standard stars.

This work aims to validate and improve the photometric precision of the Stetson standard stars using the BEST database. This paper is organized as follows. In Section\,\ref{sec:data}, we briefly introduce the latest release of the Stetson standards and BEST $UBVRI$ photometry. An independent validation and re-calibration of the Stetson photometry are presented in Sections\,\ref{sec:validation} and \ref{sec:re-calibration}, respectively. The independent validation of the re-calibrated Stetson photometry with the SCR stars and the data record of the re-calibrated Stetson catalog are described in \ref{sec:scr} and \ref{sec:record}, respectively. Finally, a summary is given in Section\,\ref{sec:summary}.

\section{Data} \label{sec:data}
\subsection{The Stetson Standard Star Database}
The Stetson standard star database \citep{Stetson2000, Stetson2005, Stetson2019} was developed to address the limited coverage of faint and high-declination sources in widely used Landolt \cite{Landolt1992} and Graham \citep{Graham1982} standard star catalogs. The first release of Stetson database was presented by \citep{Stetson2000}, based on 224 observation nights and 69 observing runs. It compiled 1,092,401 independent photometric measurements for 28,552 stars, with magnitudes ranging from 6 to 16 in the $BVRI$ bands. From this dataset, approximately 15,000 stars were selected as standard stars by requiring at least five observations obtained under photometric conditions, mean photometric uncertainties smaller than 0.02\,mag in at least two bands, and no evidence of intrinsic variability.

\begin{figure}
    \centering
    \includegraphics[width = \linewidth]{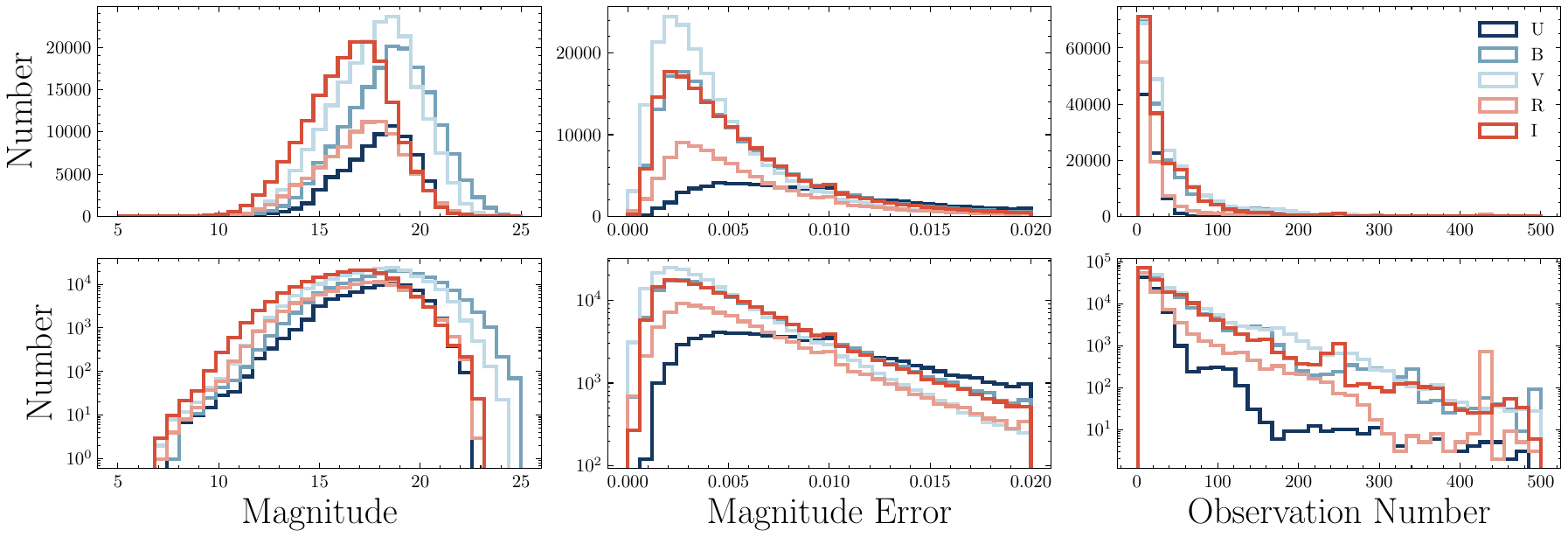}
        \caption{\small Histograms of the distributions of magnitudes, magnitude errors, and numbers of observations for all S24 sources, shown on linear (top panels) and logarithmic (bottom) scales from left to right. Colors denote different photometric bands, and the legends are shown in the upper-right corner of the top-right panel.}
        \label{fig:f1}
\end{figure}

The Stetson database has continued to update \citep{Stetson2005,Stetson2019}, steadily expanding both the number of standard stars and the volume of photometric measurements. As shown in Figure\,\ref{fig:f1}, the latest release of the Stetson 2024 standard star database (October 2024, hereafter S24) includes sources with more than 100 individual photometric observations, with the distribution peaking at $\sim$20 observations per source. The S24 database spans a wide range of stellar properties, covering magnitudes from $\sim$6 to $\sim$24 in $V$-band and $B-V$ colors from $-0.4$ to 2.4. In addition, JKC $U$-band photometry has been added for relatively bright sources, while the photometric uncertainties are maintained below 0.02\,mag.

\subsection{BEST}
The BEST database \citep{BEST} provides over 200 million XPSP standard stars with magnitudes in the Landolt $UBVRI$-bands, derived from corrected Gaia DR3 XP spectra ($G<17.65$; \cite{Carrasco2021}) in combination with the JKC transmission curves from \citet{BessellMurphy2012}. These standard stars are distributed across the entire sky and are calibrated in the AB magnitude system \citep{OkeGunn1983}. A magnitude- and color-dependent polynomial transformation is then applied to convert the Landolt standard stars in BEST to the Vega magnitude system \citep{Johnson1953, Johnson1955}, with the absolute zero-point calibrated to the Landolt standards, as mentioned in \cite{Huang2026}. 

It should be noted that although the magnitude range of the BEST standard stars does not fully overlap with that of the S24 standards, the two catalogs share a sufficient number (dozens to hundreds) of common bright stars for each filed (as illustrated in the middle panels of Figure\,\ref{fig:f2}) and no systematic magnitude-dependent errors are present in the S24 standard stars (as discussed in Section\,\ref{sec:validation}), thereby allowing for a reliable validation and re-calibration of the S24 standards using the BEST standard stars.

\subsection{Calibration Sample}
We cross-match the S24 database with the BEST–Landolt $UBVRI$ photometry constructed using the XPSP method, adopting a matching radius of $1^{\prime\prime}$. Stars are selected with $0.5 \le G_{\rm BP}-G_{\rm RP} \le 1.5$, $G \ge 10$ and \texttt{phot}\_\texttt{bp}\_\texttt{rp}\_\texttt{excess}\_\texttt{factor} $<$ $1.3+0.06\times(G_{\rm BP}-G_{\rm RP})^2$ as calibration sample for each bands. After these selections, a total of 38,256, 66,671, 73,332, 32,583, and 63,679 calibration stars are retained in the $U$, $B$, $V$, $R$, and $I$ bands, respectively. 

The spatial distribution of the number of calibration stars per field is shown in the middle panel of Figure\,\ref{fig:f2}. Most fields (71\%) contain more than 50 calibration stars, while approximately 19\% of the fields contain fewer than 25 calibration stars.

\section{Independent Validation of S24 UBVRI Photometry} \label{sec:validation}
Photometric calibration aims to characterize and correct the complex and strongly coupled systematic effects introduced by the Earth’s atmosphere and instrumental response, which can be modeled as functions of magnitude, color, and spatial position \citep{Huang2022SSPMA}. In this section, we will examine the photometry of the S24 standard stars by analyzing the consistency between the BEST and S24 measurements as a function of magnitude, color, and spatial position.

\begin{figure}
    \centering
    \includegraphics[width = \linewidth]{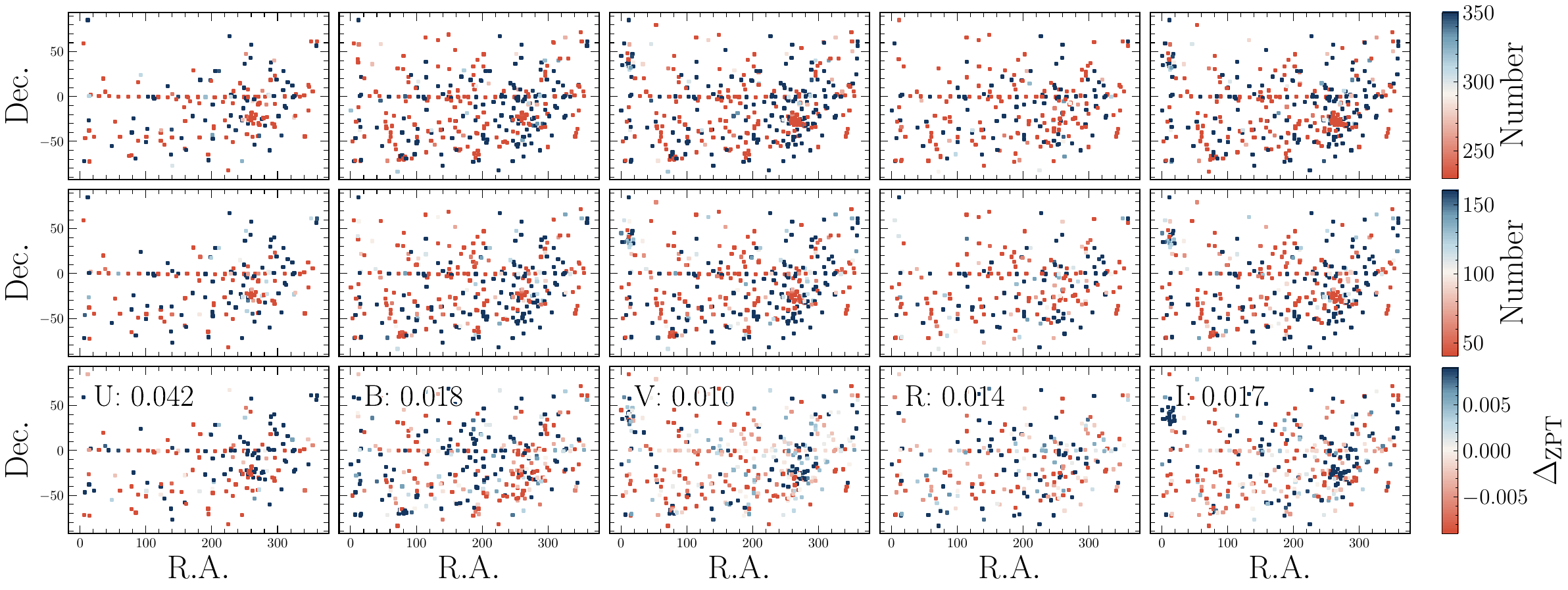}
        \caption{\small Spatial distributions of S24 source number (top), calibration star number (middle), and the zero-point offsets (bottom) for the $UBVRI$ bands from left to right. Each point denotes a S24 field, with color bars shown on the right.}
        \label{fig:f2}
\end{figure}
\begin{figure}
    \centering
    \includegraphics[width = \linewidth]{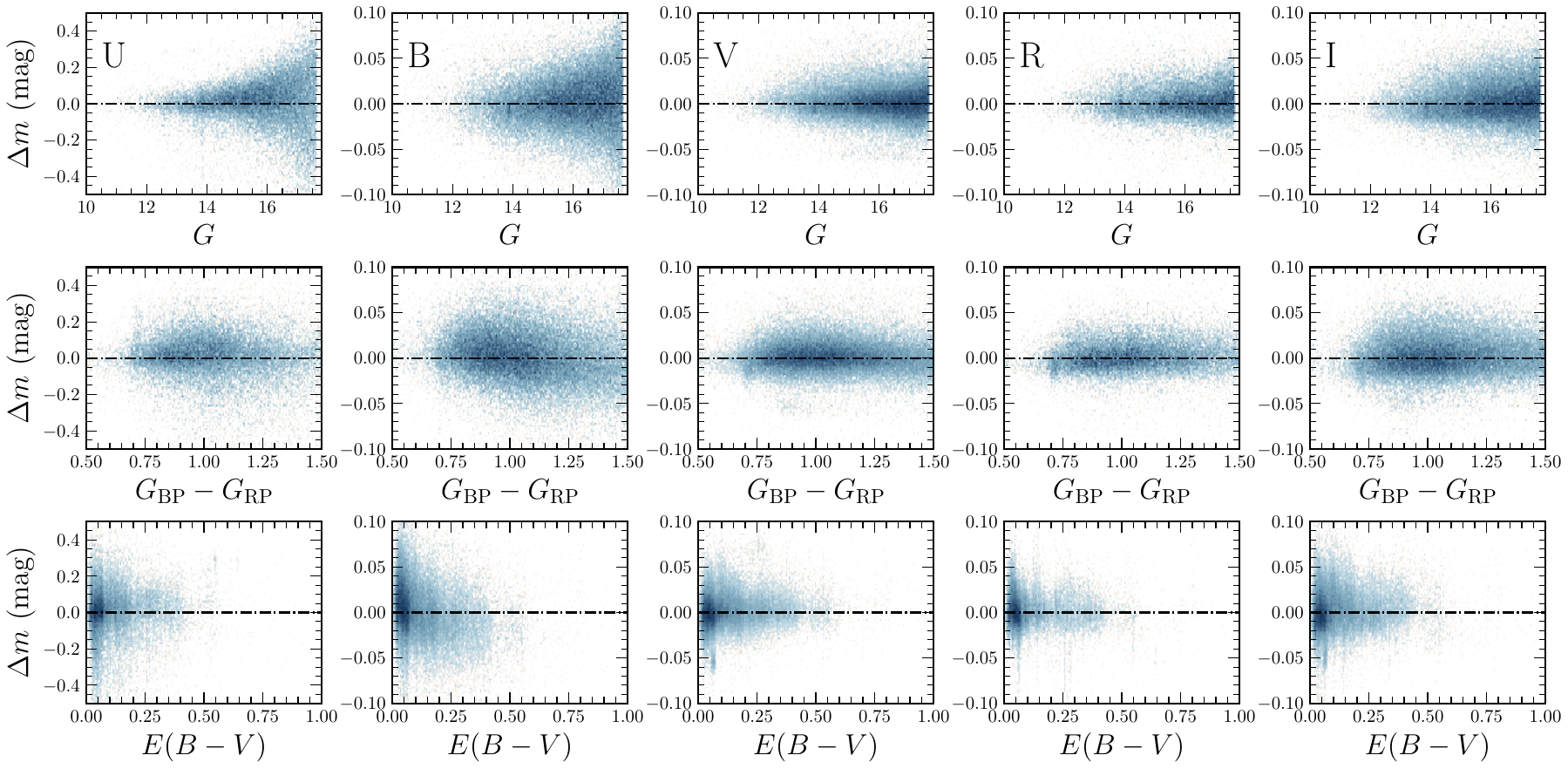}
        \caption{\small Magnitude difference between BEST and S24 as function of Gaia $G$ magnitude, Gaia $G_{\rm BP}-G_{\rm RP}$ color, and $E(B-V)$ extinction \citep{Wang2025}. Panels from left to right correspond to $U$, $B$, $V$, $R$, and $I$ bands, with colors indicating density.}
        \label{fig:f3}
\end{figure}
\begin{figure}
    \centering
    \includegraphics[width = \linewidth]{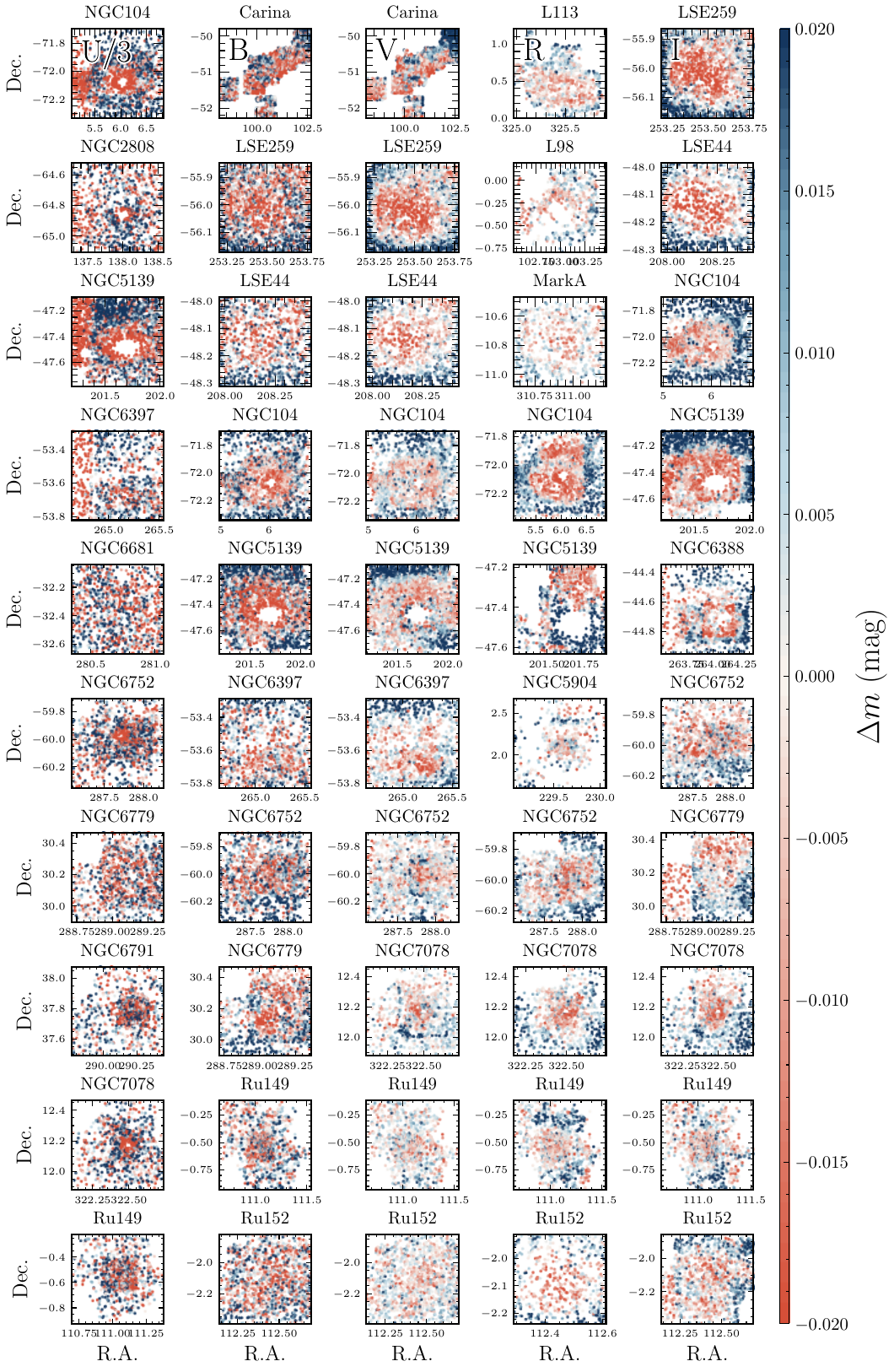}
        \caption{\small Spatial distribution of the magnitude differences between the BEST and S24 for individual fields. A color bars is shown on the right. \textcolor{black}{It should be noted that in certain fields (e.g., Ru149), the calibration stars do not cover the entire area, primarily because the S24 catalog itself does not extend to those regions.}}
        \label{fig:f4}
\end{figure}

We first examine the consistency between the S24 and BEST magnitudes as a function of magnitude $G$, color $G_{\rm BP}-G_{\rm RP}$, and extinction $E(B-V)$ from \cite{Wang2025}), as shown in Figure\,\ref{fig:f3}. The top panels show that the magnitude differences between BEST and S24 exhibit no significant trends with magnitude in any of the five bands, indicating that the CCDs used in the S24 observations exhibit good linearity and providing important evidence for re-calibrating the S24 standard stars using only relatively bright calibration samples. The middle and bottom panels further show no significant dependence on color or extinction in the $UVRI$-bands, with a weak and negligible dependence in the $B$ band. The excellent agreement between the S24 and BEST magnitudes in both magnitude and color also indicates that the two datasets are effectively on the same photometric system.

Next, we examine the consistency between the S24 and BEST magnitudes as a function of spatial position. This analysis is divided into two steps: the spatial uniformity of the photometric zero-points from field-to-field, and the uniformity of the zero points within individual fields. For the former, a constant zero-point is defined for each field as the median difference between the BEST and S24 magnitudes of the calibration stars within that field. The spatial distributions of these field-to-field zero points in the five bands are shown in the bottom panels of Figure\,\ref{fig:f2}, revealing clear spatial non-uniformity among different fields, especially for $U$-band.

To quantify this non-uniformity, we examine the histogram distributions of the field zero points and find that they are well described by Gaussian profiles. The standard deviation derived from Gaussian fitting are 0.042, 0.018, 0.010, 0.014, and 0.017\,mag in the $U$, $B$, $V$, $R$, and $I$ bands, respectively, indicating that the zero-point precision of the S24 standard star photometry is at the level of 0.01--0.04\,mag. These systematic offsets may be related to zero-point systematics in the Landolt standard stars \citep{Huang2026} and to the initial photometric calibration procedures adopted for the S24 standard stars.

For the latter, we further investigate the residuals between the BEST and S24 magnitudes as functions of position within individual fields. Prior to this analysis, the residuals in each field were corrected by subtracting the corresponding constant zero-point. We find that nearly all fields exhibit spatial variations in the magnitude offsets at a magnitude exceeding 0.01. Several representative examples are shown in Figure\,\ref{fig:f4}, where some fields display smooth, vignetting-like patterns, while others show more complex spatial structures. These systematic effects are primarily attributable to the S24 standard star data. For example, the continuous, vignetting-like variations observed within individual fields may plausibly be associated with flat-field correction in the S24 data processing. As the S24 database is constructed from multiple observations combined over time, it is not straightforward to precisely trace the origin of these systematic errors.

\section{Photometric Homogenization of Stetson Photometry} \label{sec:re-calibration}

\begin{figure}
    \centering
    \includegraphics[width = \linewidth]{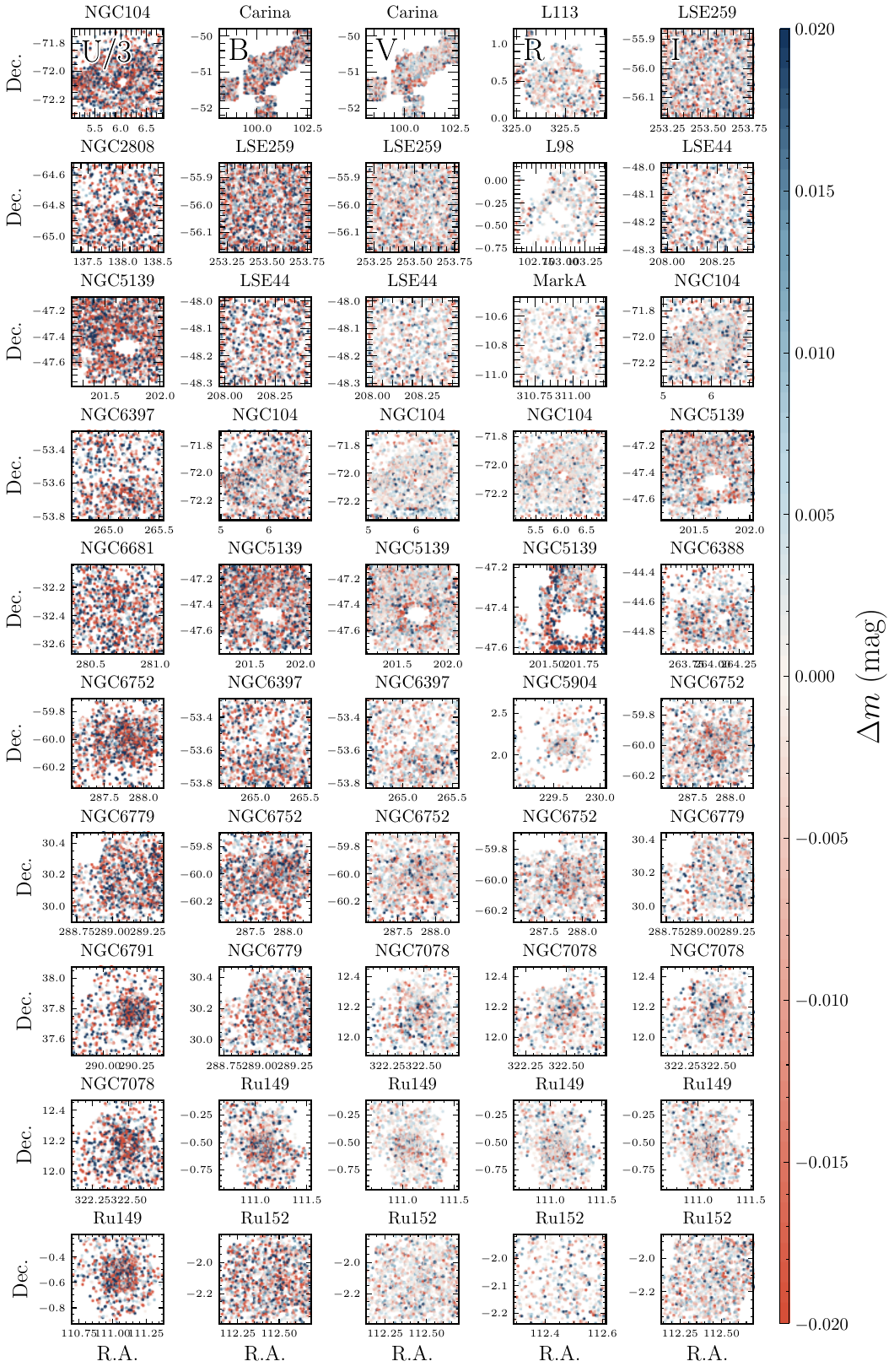}
        \caption{Same as Figure \ref{fig:f4}, but for the re-calibrated photometry.}
    \label{fig:f5}
\end{figure}
\begin{figure}
    \centering
    \includegraphics[width = \linewidth]{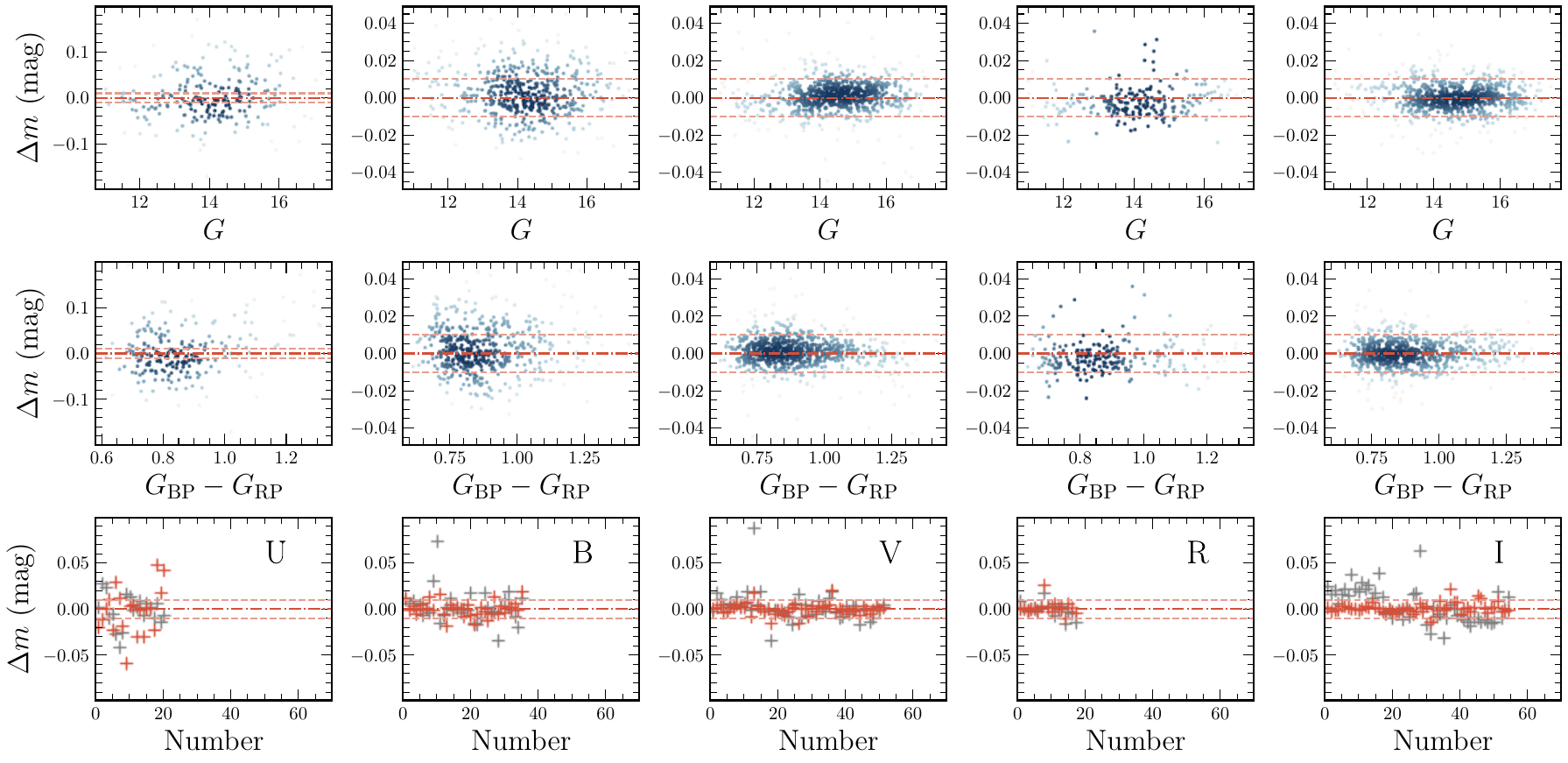}
        \caption{\small The top and middle panels show the magnitude differences as functions of the Gaia $G$ magnitude and the $G_{\rm BP}-G_{\rm RP}$ color, respectively. The bottom panels display the zero-point offsets between the SCR standards and the S24 photometry as a function of the number of calibration stars in each field, and the red (gray) points represent the corrected (original) S24 photometry. The three horizontal red lines indicate offsets of $+0.01$, 0, and $-0.01$, from top to bottom for each panel.}
    \label{fig:f6}
\end{figure}

To accurately correct the spatially correlated systematic effects present in the S24 photometry, we adopt a two-step re-calibration procedure. First, based on the results shown in the bottom panels of Figure\,\ref{fig:f2}, the field-to-field zero-point non-uniformity is corrected by applying a constant zero-point offset to all stars within each S24 field. Second, for fields containing at least 25 calibration stars, we further correct the residual intra-field spatial systematics using a numerical stellar flat correction method \citep{Xiao2024,Xiao2025MST,Li2025,Ma2025} based on the remaining magnitude offsets of the calibration stars. For fields with fewer than 25 calibration stars (accounting for approximately $\sim$19\%), only the constant zero-point correction is applied.

The numerical stellar flat procedure is performed as follows. For each S24 star within a given field, we search for the nearest 25 calibration stars in the local (RA, Dec)-space, accounting for the sky-projection effect along the right ascension direction. \textcolor{black}{The choice of 25 is adopted empirically to achieve a balance between minimizing random errors and preserving high spatial resolution.} We then fit a linear relation between the residual magnitude offsets of these calibration stars and their positions in (RA, Dec). This relation is evaluated at the position of the target star to derive its spatial-dependent magnitude correction.

The re-calibrated results of numerical stellar flat correction are presented in Figure\,\ref{fig:f5}. A significant improvement in the photometric consistency across different spatial positions within individual fields is evident after re-calibration. To quantify this improvement, we compute the standard deviation of the magnitude differences between the re-calibrated S24 and BEST photometry within each field using Gaussian fitting. For fields containing more than 50 calibration stars, the typical scatter is reduced to 5\,mmag, indicating an intra-field zero-point uniformity of better than 10\,mmag for the re-calibrated S24 standard stars, under the assumption that the photometric calibration of the XP spectra is spatially uniform.

\section{Validation of the Re-calibration of Stetson Photometry} \label{sec:scr}

This section presents independent validations of the re-calibrated S24 photometry in the five $UBVRI$ bands. In the absence of repeated measurements for individual sources in the S24 database, the internal consistency of multi-epoch observations cannot be directly used to assess the photometric calibration accuracy. Instead, the agreement between independent calibration approaches provides a feasible and reliable means to evaluate the precision of the re-calibrated magnitudes.

\cite{Huang2026} provide a catalog of more than five million Landolt-system photometric standard stars constructed using the SCR method based on LAMOST spectroscopic parameters. This catalog offers an excellent independent dataset for validating the re-calibrated S24 photometry. We cross-match the re-calibrated S24 database with the SCR standard stars and further require the Stetson standards to have more than 20 observations. We first examine the consistency between the SCR photometry and the re-calibrated S24 magnitudes. We find that the agreement is better than 0.01~mag in the $BVRI$ bands. In the $U$ band, the random uncertainties of both the SCR standards and the BEST XPSP standards are larger than those of the Stetson photometry. We further examine the dependence of the magnitude differences on stellar magnitude and color, as shown in the top and middle panels of Figure\,\ref{fig:f6}, and find no significant trends.

We also compare the field-level zero points derived from the SCR standards with those of the re-calibrated S24 database. The results are shown in the bottom panels of Figure\,\ref{fig:f6}, where a substantial improvement in the zero-point agreement is evident in the $BVRI$ bands after re-calibration. We note that the number of SCR standards per S24 field is relatively small, particularly in the $U$ and $R$ bands, where the maximum number is typically $\sim$20, limiting a robust quantitative assessment of the zero-point consistency. For the $B$, $V$, and $I$ bands, we perform Gaussian fitting to the zero-point differences for fields containing more than 20 SCR standards and obtain dispersions of 0.004, 0.002, and 0.003\,mag, respectively.

\begin{figure}
    \centering
    \includegraphics[width = \linewidth]{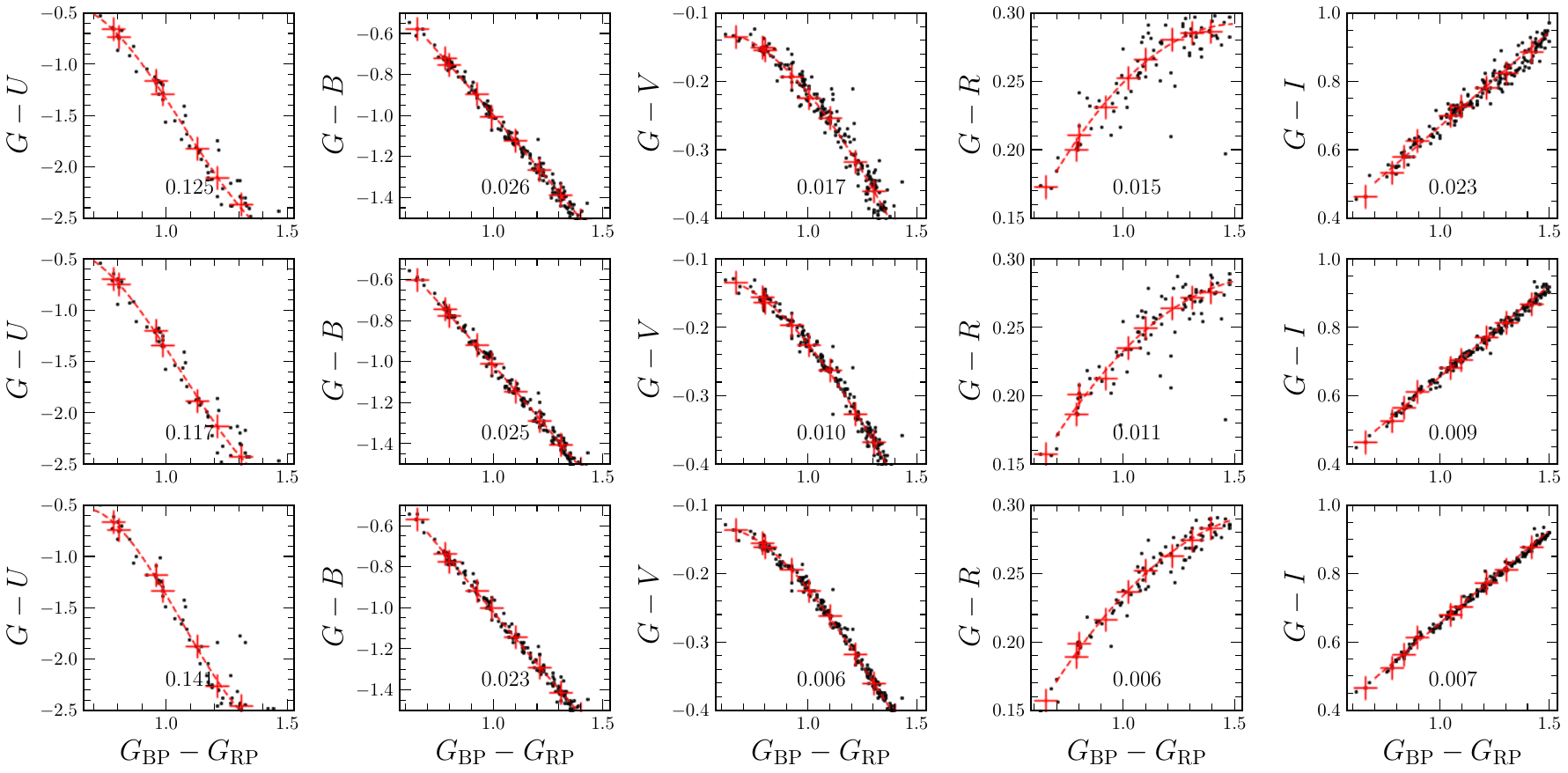}
    \caption{\small Color–color diagrams constructed using the original S24 (top), the re-calibrated S24 (middle), and the BEST photometry (bottom), each combined with Gaia three-band magnitudes. From left to right, the panels correspond to the $U$, $B$, $V$, $R$, and $I$ bands. The red lines indicate the best-fitting relations, and the corresponding fitting residuals are shown at the bottom of each panel.}
    \label{fig:f7}
\end{figure}

In addition, we assess the effectiveness of the re-calibration using Gaia three-band photometry. We cross-match the original and re-calibrated S24 databases with the XPSP standards, Gaia DR3 photometry, and the Gaia geometric distance catalog of \cite{BailerJones2021}. Using the distances and sky positions, we estimate $E(B-V)$ for each source and select a sample of more than 1000 stars with $G>15$, $E(B-V)<0.015$ and \texttt{phot}\_\texttt{bp}\_\texttt{rp}\_\texttt{excess}\_\texttt{factor} $<$ $1.3+0.06\times(G_{\rm BP}-G_{\rm RP})^2$ for this test. These sources have measurements in the original Stetson system, the re-calibrated Stetson system, the BEST photometry, and the Gaia $G$, $G_{\rm BP}$, and $G_{\rm RP}$ bands. We construct color--color diagrams using $(G_{\rm BP}-G_{\rm RP})$ versus $(G-U/B/V/R/I)$. Median points are computed in bins of $(G_{\rm BP}-G_{\rm RP})$, and the color loci are fitted with first- to third-order polynomials. The standard deviations of the residuals, derived via Gaussian fitting, are shown in Figure\,\ref{fig:f7}. Compared to the original S24, the color loci constructed from the re-calibrated magnitudes exhibit significantly reduced scatter, demonstrating the effectiveness of the re-calibration procedure. We note that the dispersion of the color loci is dominated by intrinsic physical broadening (e.g., metallicity influence, see \cite{2015ApJ...799..134Y,Yuan2015} and/or \cite{Huang2026} for details), with additional contributions from photometric random and systematic errors, and extinction. Therefore, the relative reduction in scatter is meaningful, although it should not be interpreted as a direct measure of photometric precision.

\begin{table}
    \centering
    \begin{tabular}{rclc}\toprule
         Name&  Format& Description &Unit\\\midrule
         {\tt filename}&  String& - &-\\
         {\tt starname}&  String& - &-\\ 
 {\tt ra}& Float&Right Ascension (J2000) &deg ($^\circ$)\\
 {\tt dec}& Float&Declination (J2000) &deg ($^\circ$)\\
 {\tt U}& Float&S24 $U$ band magnitude  &mag\\
 {\tt eU}& Float&S24 $U$ band magnitude error  &mag\\
 {\tt NU\_1}& \textcolor{black}{Integer}& Number of $U$ band observations &-\\
 {\tt nU\_1a}& \textcolor{black}{Integer}& Number of $U$ band photometric observations &-\\
 {\tt B}& Float&S24 $B$ band magnitude  &mag\\
 {\tt eB}& Float&S24 $B$ band magnitude error  &mag\\
 {\tt NB\_1}& \textcolor{black}{Integer}& Number of $B$ band observations &-\\
 {\tt nB\_1a}& \textcolor{black}{Integer}& Number of $B$ band photometric observations &-\\
 {\tt V}& Float&S24 $V$ band magnitude  &mag\\
 {\tt eV}& Float&S24 $V$ band magnitude error  &mag\\
 {\tt NV\_1}& \textcolor{black}{Integer}& Number of $V$ band observations &-\\
 {\tt nV\_1a}& \textcolor{black}{Integer}& Number of $V$ band photometric observations &-\\
 {\tt R}& Float&S24l $R$ band magnitude  &mag\\
 {\tt eR}& Float&S24 $R$ band magnitude error  &mag\\
 {\tt NR\_1}& \textcolor{black}{Integer}& Number of $R$ band observations &-\\
 {\tt nR\_1a}& \textcolor{black}{Integer}& Number of $R$ band photometric observations &-\\
 {\tt I}& Float&S24 $I$ band magnitude  &mag\\
 {\tt eI}& Float&S24 $I$ band magnitude error  &mag\\
 {\tt NI\_1}& \textcolor{black}{Integer}& Number of $I$ band observations &-\\
 {\tt nI\_1a}& \textcolor{black}{Integer}& Number of $I$ band photometric observations &-\\
 {\tt ebv}& Float& $E(B-V)$ from \cite{Wang2025}&mag\\
 {\tt rgeo}& Float& Bailer-Jones {\tt rgeo} distance&parsec\\
 {\tt U\_recali}& Float& Re-calibrated $U$ magnitude in this work &mag\\
 {\tt U\_flag}& \textcolor{black}{Integer}& $U$-band photometric quality provided in this work &-\\ 
 {\tt B\_recali}& Float& Re-calibrated $B$ magnitude in this work &mag\\
 {\tt B\_flag}& \textcolor{black}{Integer}& $B$-band photometric quality provided in this work &-\\ 
 {\tt V\_recali}& Float& Re-calibrated $V$ magnitude in this work &mag\\
 {\tt V\_flag}& \textcolor{black}{Integer}& $V$-band photometric quality provided in this work &-\\ 
 {\tt R\_recali}& Float& Re-calibrated $R$ magnitude in this work &mag\\
 {\tt R\_flag}& \textcolor{black}{Integer}& $R$-band photometric quality provided in this work &-\\ 
 {\tt I\_recali}& Float& Re-calibrated $I$ magnitude in this work &mag\\
 {\tt I\_flag}& \textcolor{black}{Integer}& $I$-band photometric quality provided in this work &-\\ 
 \bottomrule 
    \end{tabular}
    \caption{Each column of re-calibrated S24 catalog.}
    \label{tab:tableheader}
\end{table}

Finally, we compare the re-calibrated S24 photometry with the BEST XPSP magnitudes. For bright sources with $G<15$, the BEST photometry exhibits higher precision than the Stetson magnitudes. For sources with $15<G<17.65$, the re-calibrated S24 $U$-band magnitudes show smaller random uncertainties than the BEST photometry, while the two are comparable in the $B$ band, and the BEST photometry is significantly more precise in the $VRI$ bands. These results suggest that, if feasible, applying the BEST database and method presented in this work to future releases of the Stetson standard stars could substantially enhance its precision. Moreover, with the forthcoming release of Gaia DR4, a further improvement in the precision of the BEST $U$-band photometry may be expected.

\section{Data record} \label{sec:record}
The corrected S24 catalog contains all photometric sources from the S24 database, including original S24 magnitudes, magnitude errors, observation numbers, and photometric observation numbers used for calibration. It also includes the final corrected S24 magnitudes, $1''$ cross-matched Bailer-Jones distances {\tt rgeo} \citep{BailerJones2021} and corresponding $E(B-V)$ \citep{Wang2025}. For calibrated S24 photometric results, we use flags to help determine data validity: calibration stars within a single Stetson field fewer than 25, or stars with an invalid S24 photometry, are marked as {\tt 0}. All other values are marked as {\tt 1}. The final catalog is a FITS format file of 40.3 MB size, comprising 36 columns. Column names, data formats, descriptions, and units are listed in Table \ref{tab:tableheader}. \textcolor{black}{The recalibrated Stetson catalog is publicly available at https://nadc.china-vo.org/res/r101783/ and can also be accessed via the BEST website \footnote{https://nadc.china-vo.org/data/best/}.}

\section{Conclusions} \label{sec:summary}

In this paper, we perform an independent validation and re-calibration of the S24 standard star photometry using the BEST database. Using a total of typically 30,000-70,000 standard stars per band, we show that the zero-point precision of the original S24 photometric calibration is approximately 42, 18, 10, 14, and 17\,mmag in the $U$, $B$, $V$, $R$, and $I$ bands, respectively. In addition, significant spatial variations in the magnitude offsets are detected within individual Stetson fields for all five bands, reaching magnitudes exceeding 1\%. These variations are attributed to spatially dependent calibration errors in the original Stetson photometry.

To correct these effects, we first remove field-to-field zero-point inhomogeneities by applying constant zero-point offsets derived from the median differences between the BEST and Stetson magnitudes for calibration stars in each field. We then correct the residual intra-field spatial systematics using a numerical stellar flat-fielding approach. After re-calibration, the consistency between the Stetson and BEST magnitudes is improved to a level of $\sim$5\,mmag. The re-calibrated Stetson photometric catalog has been publicly released.

We further compare the re-calibrated Stetson photometry with the SCR standards constructed by \citep{Huang2026} and find an agreement better than 10\,mmag in the $BVRI$ bands, with no detectable dependence on magnitude or color. This comparison independently confirms a zero-point precision of approximately 2--4\,mmag for the re-calibrated Stetson photometry in the $BVI$ bands. In addition, the improvement in the Stetson photometry achieved in this work is qualitatively validated using Gaia DR3 three-band photometry.

Finally, using Gaia DR3 photometry, we compare the re-calibrated Stetson magnitudes with the BEST $UBVRI$ magnitudes derived from Gaia XP spectra. For sources with $15 < G < 17.76$, we find that the re-calibrated Stetson photometry exhibits higher overall precision (including both random and systematic components) than BEST in the $U$ band, comparable precision in the $B$ band, and lower in other bands. Moreover, with the forthcoming release of Gaia DR4, further improvements in the precision of the BEST $U$-band photometry may reasonably be expected.

These results demonstrate the power of the BEST database in improving the calibration precision of wide-field photometric surveys. We suggest that, if feasible, the BEST database be incorporated into the calibration process of future releases of the Stetson standard star database.

\clearpage
\section*{acknowledgments}
\textcolor{black}{We thank the anonymous referee for the helpful comments.}
This work is supported by the National Natural Science Foundation of China grants No. 12403024, 12422303, 12222301, 12173007, 12273077, 12573033, and 124B2055; the National Key R\&D Program of China (grants Nos. 2023YFA1608303, SQ2024YFA160006901, 2024YFA1611601, and 2022YFF0711500); the Postdoctoral Fellowship Program of CPSF under Grant Number GZB20240731; the Young Data Scientist Project of the National Astronomical Data Center; the China Postdoctoral Science Foundation No. 2023M743447; and the Natural Science Foundation of Shandong Province (Grant No.ZR2025MS81 and No.ZR2022MA076). 

\clearpage
\bibliographystyle{raa}
\bibliography{bibtex}

\end{document}